\documentclass{article}

\usepackage{arxiv}

\usepackage[utf8]{inputenc} 
\usepackage[T1]{fontenc}    
\usepackage{hyperref}       
\usepackage{url}            
\usepackage{booktabs}       
\usepackage{amsfonts}       
\usepackage{nicefrac}       
\usepackage{microtype}      
\usepackage{lipsum}
\usepackage[numbers]{natbib}
\usepackage{lineno,hyperref}
\usepackage{booktabs} 
\usepackage{colortbl}
\usepackage{multirow}
\usepackage{amssymb}
\usepackage{hyperref}
\usepackage{mathtools}
\usepackage{tikz}
\usepackage[flushleft]{threeparttable}
\usepackage{times,amsmath,epsfig}

\title{Benchmarking Graph Data Management and Processing Systems: A Survey}

\author{
  Miyuru Dayarathna\\
  WSO2, Inc.\\
  3080 Olcott St, Suite C220,\\
  Santa Clara, CA, 95054, USA\\
  \texttt{miyurud@wso2.com} \\
   \And
 Toyotaro Suzumura \\
 The University of Tokyo\\
 Tokyo, Japan\\
  \texttt{suzumura@acm.org} \\
}

\begin{document}
\maketitle

\begin{abstract}
The development of scalable, representative, and widely adopted benchmarks for graph data systems have been a question for which answers has been sought for decades. We conduct an in-depth study of the existing literature on benchmarks for graph data management and processing, covering 20 different benchmarks developed during the last 15 years. We categorize the benchmarks into three areas focusing on benchmarks for graph processing systems, graph database benchmarks, and bigdata benchmarks with graph processing workloads. This systematic approach allows us to identify multiple issues existing in this area, including i) few benchmarks exist which can produce high workload scenarios, ii) no significant work done on benchmarking graph stream processing as well as graph based machine learning, iii) benchmarks tend to use conventional metrics despite new meaningful metrics have been around for years, iv) increasing number of big data benchmarks appear with graph processing workloads. Following these observations, we conclude the survey by describing key challenges for future research on graph data systems benchmarking.
\end{abstract}

\keywords{System Performance \and Benchmarking \and Graph Databases \and Big Data \and Large Graph Analytics}

\section{Introduction}
\label{sec:introduction}
Graph data related use cases are frequently found in recent computing applications \cite{6597145}. Both academia and industry have developed number of systems to address the requirement of analyzing graph data sets. There are several key features which distinguishes the handling of graph data from other data storage and analysis scenarios. First, graph data are connected data, where storage and processing of such data need to consider the connections between them. Such graph data sets need to be traversed frequently. Another characteristics of graph data is they do not explicitly enforce a schema on the data even when the data may be implicitly structured \cite{Aluc:2014:WMW:2732951.2732957}.

Currently, there are two key paradigms for large graph data analysis called batch graph data analysis and online graph data analysis \cite{Dayarathna:2014:TSD:2689676.2689677}\cite{ref2:Shao:2013:TDG:2463676.2467799}\cite{ref1:Shao:2012:MML:2213836.2213907}. In the batch graph data analysis the graph data set is loaded from a storage media as a whole into the graph processing application and the graph algorithms are run on top of the loaded data set. Pegasus \cite{Kang2014}, Hama \cite{7752866}, Giraph \cite{Ching:2015:OTE:2824032.2824077}, etc. are examples for systems developed following MapReduce paradigm. The graph algorithms conduct resource intensive analysis on the graph data sets. In the online transaction processing of graph data sets, graphs are loaded from the storage media into a graph processing system and multiple types of queries are executed on the loaded graph data sets. This should not be confused with online graph analytics which operate on dynamic graphs which appear as data streams. Systems such as graph databases, Graph based recommender systems \cite{Sharma:2016:GRC:3007263.3007267}, etc. fall into this category.

Users of graph processing platforms often face the requirement of exploring the performance of graph-processing platforms and their specific applications \cite{Guo:2014:WGP:2650283.2650530}. Key performance factors include data sources, frameworks, computation types, and data representations \cite{6529300}. Some of the frameworks are better in executing some algorithms. Data sources, the data representations of these sources plays important role in determining the system performance. Due to these reasons it becomes very difficult for an end-user to determine which framework best matches for solving a problem from multiple different angles such as performance, productivity, economic benefits, etc. \cite{Satish:2014:NMG:2588555.2610518}.

The performance of the graph processing systems are measured using domain-centric benchmarks especially targeted for evaluating the graph data processing capabilities of such systems \cite{10.1007/978-3-642-36727-4_8}. There have been plethora of work conducted on performance analysis and benchmarking of graph data management and processing systems \cite{Dayarathna2017}. However, most of these efforts are scattered in multiple research efforts and the research field lacks detailed taxonomy. The goal of this survey is to fill this void by creating a detailed yet coherent taxonomy of the existing work on graph data management and processing system performance analysis. It is quite common fact that in most of the research conducted on graph data processing systems the evaluation sections of the research papers consist of some kind of performance analysis with the competitive systems. It is because system performance improvement has become the best approach for showcasing the importance of the expressed improvement over the state-of-the-art. However, in this survey we avoid taking such research results into account as much as possible and we focus on the studies conducted purely with the motivation of performance comparison of systems. Perhaps this could be a separate theme for a survey on graph data processing performance. However, we incorporate certain important performance results which are required for explaining phenomenons of graph data processing system behavior in this survey from such papers as well. Note that all the work listed in the subsections of this paper (unless specified) are chronologically ordered from oldest to the latest.

We start the paper by discussing the related surveys conducted in this area. Through this we emphasize the need for a comprehensive survey for graph data processing and management system performance. Next, we list the leading performance improvement efforts on graph database systems. Furthermore, most of the existing benchmarks are targeting narrow use cases whereas multiple new applications of graph data processing has appeared recently. This has brought forward the need for developing sophisticated benchmarks for graph data processing. Next, we present a process for design and implementation of graph processing system benchmarks. We show how this general process can be applied to specific benchmarks by taking examples from literature. Finally, we describe key research challenges for future research on performance of graph data processing systems. In the next section we present related surveys for our work.

\section{Related Surveys}
\label{sec:related-surveys}
Multiple research have been done on graph data processing and management systems as well as their benchmarking. However, not much work have been conducted to survey these areas. We classify the related surveys published to-date into two classes as surveys on graph databases and surveys on graph processing systems.

\subsection{Surveys on Graph Databases}
\label{subsec:related-surveys-gdb}
Surveys conducted on graph databases list the existing leading graph database development efforts. Angles \emph{et al.} did one of the significant research in this context which surveyed the graph database models in detail \cite{Angles:2008:SGD:1322432.1322433}. They provided definition for graph database models by considering the implicit and informal notions used in the literature. They created their presentation on the main aspects of graph data modeling such as data structures, query languages, integrity constraints, etc. They also made a comparison of graph database models with respect to other database models. Furthermore, they provided a uniform description of the most relevant graph database models.

\subsection{Surveys on Graph Processing Systems}
\label{subsec:related-surveys-graph-systems}
Multiple surveys have been conducted on graph processing system models. Majority of them were conducted in related to graph processing programming models. A survey on parallel graph processing frameworks was made by Doekemeijer \emph{et al.} \cite{Doekemeijer:2014a}. They developed a taxonomy of more than 80 graph processing systems which are aimed at efficient processing of large graphs. Similarly McCune \emph{et al.} surveyed the work conducted on vertex-centric approach for graph processing \cite{McCune:2015:TLV:2830539.2818185}. They investigated on ``Think Like a Vertex'' (TLAV) frameworks focusing on the concepts and system architectures. The survey mainly investigates on the four principles of which the TLAV frameworks have been developed called Timing, Communication, Model of Computation, and Partitioning. McGregor created a survey of graph stream algorithms \cite{McGregor:2014:GSA:2627692.2627694}. They discussed about some important summary data structure such as spanners and sparsifiers. The survey mainly covers areas such as insert-only streams, graph sketches, and sliding windows. Batarfi \emph{et al.} did a survey of the cutting edge large graph processing platforms \cite{Batarfi2015}. The study was conducted on five famous graph processing systems. Similarly Junghanns \emph{et al.} surveyed about current techniques for handling and processing of large graphs \cite{Junghanns2017}. Yan \emph{et al.} made a categorization of large graph processing systems \cite{DBS-056}. 

There are few surveys conducted on graph processing on-non traditional hardware. Tran \emph{et al.} reviewed the work done on adapting massively parallel architecture of Graphical Processing Units (GPUs) to improve the performance of basic graph operations \cite{Tran2018}. They found that architecture related issues of GPUs, issues of programming models, and structure related issues of graphs as the hindrance for achieving high performance with GPU. Similarly Shi \emph{et al.} conducted a survey on main concerns of GPU based graph processing \cite{shi2018graph}. Heidari \emph{et al.} developed a graph analysis system classification \cite{Heidari:2018:SGP:3212709.3199523}. However, all these surveys are different from ours because we specifically focus on performance benchmarks of both graph data management and processing systems. 

Surveys conducted on graph processing performance are quite rare. Bonifati \emph{et al.} described the features of graph processing benchmarks \cite{Bonifati2018}. They studied about distributed and parallel benchmarks for graph analyzers, graph database benchmarks, benchmarks for RDF databases, and data-only benchmarks. Their work was primarily on benchmarks while our work goes beyond them. We not only describe the benchmarks, but also describe wide variety of aspects such as benchmark metrics, workload characterizations conducted using benchmarks, etc. which are very important aspects of system performance.

\begin{table*}[htbp]
\centering
\begin{center}
	  	\caption{Related surveys.}
  		\label{table:related-surveys}
\begin{tabular}{ | p{0.7cm} | p{4cm} | p{10cm} |}
\hline
Year & Investigator(s) & Area of study \\ \hline
2008 & Angles \emph{et al.} \cite{Angles:2008:SGD:1322432.1322433} & Graph database models \\ \hline
2014 & Doekemeijer \emph{et al.} \cite{Doekemeijer:2014a} & Parallel and distributed large graph processing frameworks \\ \hline
2014 & McGregor
\cite{McGregor:2014:GSA:2627692.2627694} & Algorithms for graph stream analysis \\ \hline
2015 & McCune \emph{et al.} \cite{McCune:2015:TLV:2830539.2818185} & Vertex-centric graph processing frameworks \\ \hline
2015 & Batarfi \emph{et al.} \cite{Batarfi2015} & Systems for processing large graphs \\ \hline
2017 & Junghanns \emph{et al.} \cite{Junghanns2017} & Approaches for analysis and management of large graph data \\ \hline
2017 & Yan \emph{et al.} \cite{DBS-056} & Systems for processing large graphs \\ \hline
2017 & Sahu \emph{et al.} \cite{Sahu:2017:ULG:3186728.3164139} & Systems for processing large graphs \\ \hline
2018 & Bonifati \emph{et al.} \cite{Bonifati2018} & Benchmarks for graph processing systems \\ \hline
2018 & Shi \emph{et al.} \cite{shi2018graph} & A survey of graph processing on graphics processing units (GPUs) \\ \hline
2018 & Tran \emph{et al.} \cite{Tran2018} & A survey of graph processing on GPUs \\ \hline
2018 & Heidari \emph{et al.} \cite{Heidari:2018:SGP:3212709.3199523} & Systems for processing large graphs \\ \hline
2019 & Lu \emph{et al.} \cite{10.1145/3323214} & Multi-model databases \\ \hline
2019 & Drobyshevskiy \emph{et al.} \cite{10.1145/3369782} & Random Graph Models \\ \hline
\end{tabular}
\end{center}
\end{table*}

Considerably different way for surveying graph processor related research was presented by Sahu \emph{et al.} \cite{Sahu:2017:ULG:3186728.3164139}. They conducted an online survey with the aim of understanding the graph computations users execute, the type of the graphs users process, the significant issues users encounter while analyzing the graphs, and different types of graph software users use. Their work does not focus on benchmarks.

Recently multi-model databases has become an emerging field of interest. These databases can be used to manage graph structured data. Lu \emph{et al.} provide a classification of multi-model databases \cite{10.1145/3323214}. Graph data generators are critical for implementing successful graph benchmarks. Drobyshevskiy \emph{et al.} made a comprehensive survey on random graph generators \cite{10.1145/3369782}. While these surveys are very relevant for the area surveyed by us we specifically investigate on benchmarks and their features.

A summary of the related surveys is shown in Table \ref{table:related-surveys}. In a nutshell it is apparent that there is a rising number of surveys conducted on the graph data management and processing area. However, there exists a significant void on the detailed survey conducted on graph processing benchmarks. This brings up the motivation for conducting this survey. Next, we describe a general framework on graph processing and management systems which can be used to compare features across such systems.

\section{General Framework for Graph Data Management and Processing}
\label{sec:framework}
Diverse field of graph data management and processing can be abstracted into three main areas as \emph{applications}, \emph{computation}, and \emph{storage}. Applications are the realization of graph processing use cases. There are multiple application domains which handle large graph data. Computation corresponds to the programming models and algorithms developed to solve the issue of efficient processing of large graph data. Storage is the methodologies used for handling different types of graphs and storing them to support fast access.

\begin{figure}[htbp]
	\centering
		\includegraphics[width=0.7\textwidth]{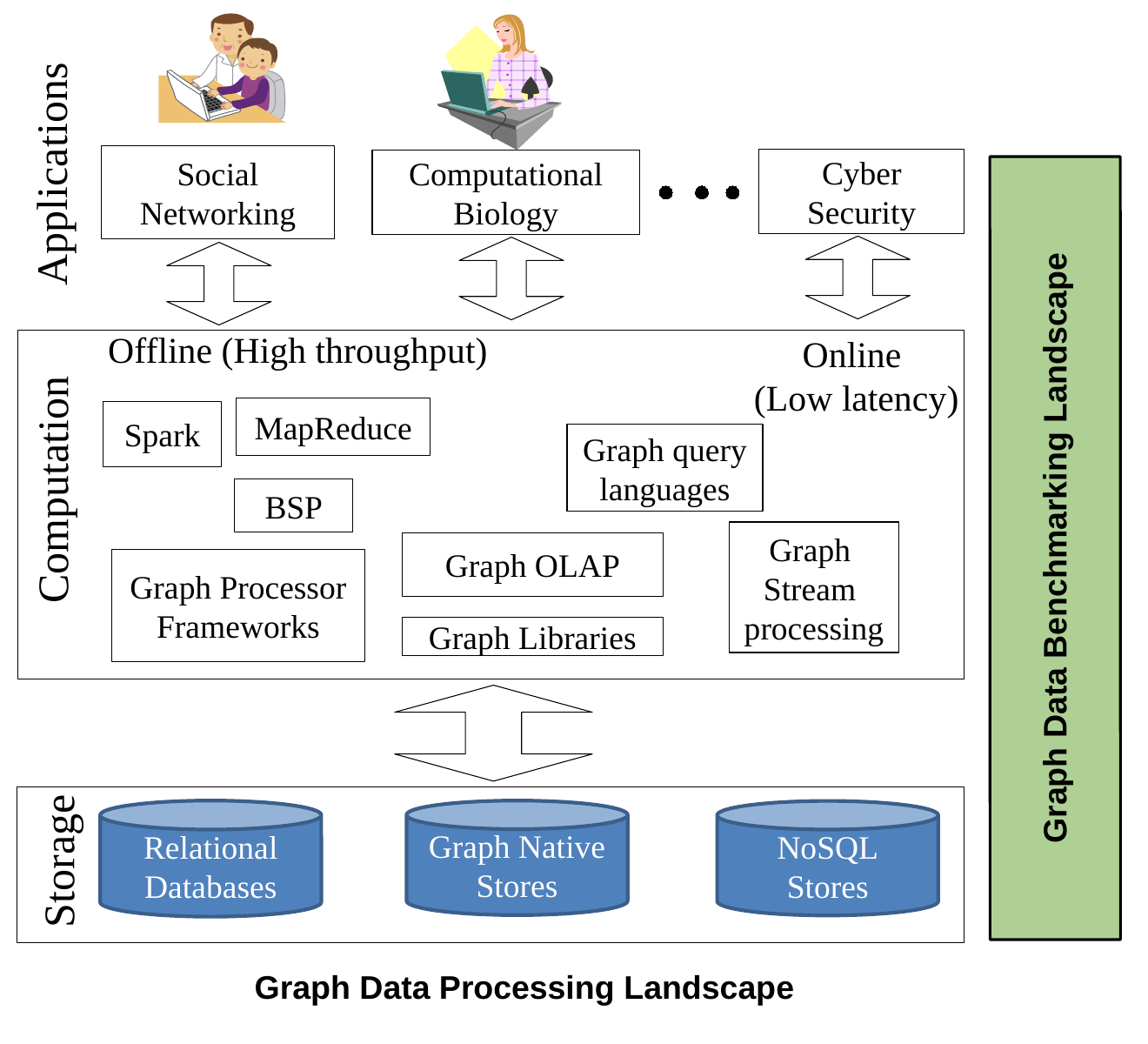}
	\caption{Overview of graph processing landscape.}
	\label{fig:graphmgtmine}
\end{figure}

There are multiple applications which operate on graph data. Some of the key areas highlighted in Figure \ref{fig:graphmgtmine} are Semantic Web, Social Networking, Computational Biology, Geographic Information Systems (GIS), and Cyber Security. These applications may operate on different graph data sets such as road networks, cyber security graphs, Facebook graph, etc. 

Graph computation on the other hand can be divide into two sub areas based on the style of processing being conducted as offline and online graph analytics. Benchmarking is a process which spans across the three main areas.

\subsection{Offline Graph Analytics}
\label{subsec:offline-framework}
The first category, offline graph analytics treats graph data processing tasks as batch jobs. For example, Pegasus graph processing system (i.e., graph processor framework) \cite{Kang2014} conducts PageRank computation on a graph that is kept on Hadoop Distributed File System (HDFS). Graph processing libraries such as PBGL \cite{Gregor:2005:LSG:1094811.1094844}, ScaleGraph \cite{6507498}\cite{Dayarathna:2012:ISX:2246056.2246062}, GraphLab \cite{7474329}, Giraph \cite{Ching:2015:OTE:2824032.2824077}, etc. are similar examples for offline graph analytics.

\subsection{Online Graph Analytics}
\label{subsec:online-framework}
The second category, online graph analytics operate on dynamic graphs where the graph structure itself changes due to updates. For example, given a vertex, a graph query can be written to identify all the neighbouring vertices. Such operations tend to execute on a graph data set multiple times. Hence, the response latency needs to be kept minimum in such applications. Graph databases such as Acacia \cite{7820312}\cite{Dayarathna2017}, Neo4j \cite{neo4jbook:2012a}, OrientDB \cite{website:orientdb:2017a}, Titan \cite{website:aurelius:2017a}, TigerGraph \cite{website:tigergraph:2018a}, etc. are examples for such online graph processing systems. Furthermore, in certain applications graph data may be received by the application as a stream of edges \cite{8287755}. For example, in an online social networking application friendships which arrive as edges need to be reflected in the graph data structure which is used to calculate the highly connected component of the social network \cite{Jayasinghe:2016:CAG:2933267.2933508}. Similarly multiple graph data processing algorithms have been implemented as streaming algorithms. Examples include Streaming PageRank \cite{Sarma:2011:EPG:1970392.1970397}, Streaming K-Core Decomposition \cite{Sariyuce:2013:SAK:2536336.2536344}, Streaming Triangle Counting \cite{Pavan:2013:CST:2556549.2556569}, etc.     

Characteristics of graph storage is mainly determined by the characteristics of the data structures used and the physical organization of the storage media. Data structures such as Resilient Distributed Data Sets (RDD) have been used to create data structures which could efficiently store graph data. In terms of physical organization memory hierarchy for example play critical role in determining the full system performance of a graph processing application. Graph stores such as Neo4j \cite{graphdatabsesbook:2013a}, Acacia \cite{8287755}\cite{Dayarathna:2014:TSD:2689676.2689677}, TigerGraph \cite{website:tigergraph:2018a}, etc. have introduced native storage structures for graph data storage as a solution. Benchmarking process conducted on such systems is described in next section.

\section{Benchmarking Process for Graph Data Management and Processing Systems}
\label{sec:benchmarking-process-for-graph-data-processing-systems}
\begin{figure}[htbp]
	\centering
		\includegraphics[width=0.7\textwidth]{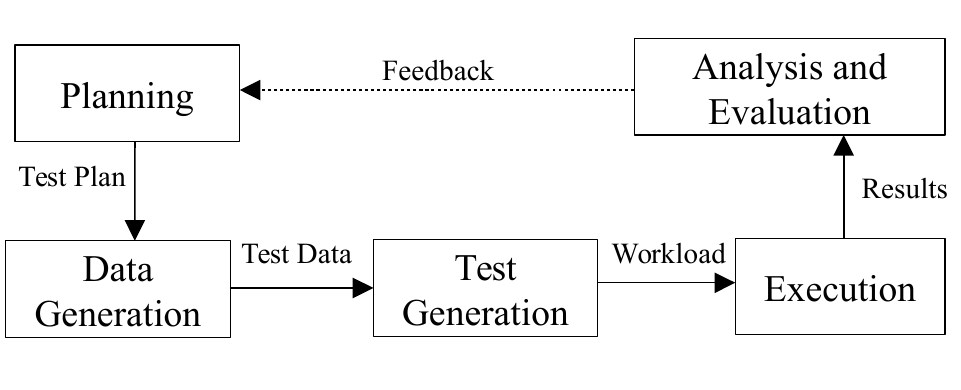}
	\caption{The process of graph data processing system benchmarking.}
	\label{fig:overview}
\end{figure}

Big data applications typically ingest and filter huge amounts of data and then conduct complex analytics on them. The designers of big data benchmarks make sure their benchmarks represent a wide range of such applications \cite{Taft:2014:GCA:2588555.2595633}. The general flow involved in benchmarking process of a graph data processing system (similar to big data benchmarking \cite{han2014big}) can be shown as in Figure \ref{fig:overview}. The entire benchmarking process can be divided into five stages as \emph{Planning}, \emph{Data Generation}, \emph{Test Generation}, \emph{Execution}, and \emph{Analysis and Evaluation} following the famous MAPE loop (Monitoring, Analysis, Planning, Execution) \cite{computing2006architectural}. MAPE was originally introduced by IBM with their architecture for autonomic computing.

The Planning stage determines the characteristics of the workload, the types of the systems to compare with, and the execution environment properties. The characteristics of the workload is very important parameter that governs the end result of the benchmarking process. Types of the systems to compare with is very important because of the existence of large variety of similar systems. Comparing graph processing systems such as igraph \cite{igraph:2019a} or SNAP (Stanford Network Analysis Project) \cite{leskovec2016snap} that are intended to run on single system against a distributed graph processing system such as Apache Giraph or Apache Spark GraphX is very difficult task. Systems should share some common features such as distributed processing, in-memory systems, programming model, etc. in order for them to be compared in a fair manner. Similarly the execution environment on which the graph processing software runs plays key role in determining their performance. Hence equivalent systems need to be used in benchmark comparisons.

The data generation stage is concerned with construction of test datasets. Important aspects of data generators include generating realistic data within short period of time, repeatability of the characteristics, scalability, etc. The data should be realistic which represents the practical scenarios. Data from similar set of parameters should possess similar characteristics. Furthermore, the data generator should be scalable in such a way that different levels of workloads can be imposed on the target system.

Test generation is the process of constructing the test case scenario. This is much more broader compared to the data generation step. Test generation involves preparation of the testing environment. For example, this step is concerned of the amount of resources such as RAM, CPU cores allocated on a computer used for deploying the target system. For distributed graph analytics the number of compute nodes on which the data processing system is deployed can also be considered as a parameter.

Execution is the step where the actual execution of the benchmark setup is conducted. The benchmarking framework invokes the planned test on the graph data processing system. Execution step can range from few minutes to multiple hours depending on the characteristics of the data processing system, workload size, as well as the test scenario. When running the benchmark multiple performance metrics are collected.

The final crucial step of this process is the analysis of the performance results. The raw performance information gathered are analyzed intelligently. In the case of a distributed graph processing system, performance characteristics of each node need to be taken into consideration since the resource intensive operation can be located in either of the nodes or in a node combination. The performance characteristics are expressed in terms of performance metrics which are discussed next.

\section{Performance Metrics for Graph Data}
\label{sec:metrics-graph-data}
The performance metrics are the values used for measuring the performance behavior of a system. The metrics play a key role in identifying the success of a performance study because the correct combination of metrics has to be utilized for doing a meaningful comparison. Variety of performance metrics are found at multiple stages of the operation of the graph data processing system. For example, a metric such as latency can be measured at graph data loading time as well as during the processing of the graph data set. Latency can be expressed as per-operation metric as well as for the complete process. In the latter case it is referred to as the elapsed time. 

Metrics for graph data processing include the number of edges traversed per second (TEPS), Latency of query execution, etc. Traversed edges per second (TEPS) is relatively new metric originated by the HPCS benchmark \cite{Dominguez-Sal:2010:SGD:1927585.1927590} and the Graph500 benchmarking initiative. TEPS is a type of throughput metric. Hence in this paper we categorize TEPS as a subcategory under throughput. Systems such as Acacia measure the query execution time (Latency) for each query it executes and prints the execution statistics on the client console. Memory foot print is also very significant performance metric for graph data processing system. Memory foot print corresponds to the amount of memory the graph processing application uses. Similarly CPU usage and network read/write are metrics used for measuring whether the system has enough CPU resources to conduct the processing as well as to check whether there is any communication bottlenecks in the system.

\begin{table*}[htbp]
\begin{center}
	  		\caption{Key Performance Metrics for Graph Data Processing System benchmarks.}
  		\label{table:metrics-for-graph-data}
\begin{tabular}{ | p{3cm} | p{3cm} | p{3cm} | p{1.6cm} | p{1.3cm} | p{1.8cm} |}
\hline
Metric & Phase of execution & Capabilities measured & Layer of application & Main users & Specific area of focus \\ \hline
Latency (L) & data processing, data loading & computational power & applications & N/A & Graph stream processing \\ \hline
Throughput (T) & data processing, data loading & computational power & applications & Graph500 & Batch graph processing \\ \hline
Memory Footprint (M) & data processing, data loading & efficient use of system memory & Operating System & N/A & N/A \\ \hline
CPU Usage (CPU) & data processing, data loading & efficient use of CPU cycles & Operating System & N/A & N/A \\ \hline
Network read/write (N) & data processing & bottleneck free communication & Operating System & N/A & N/A \\ \hline
\end{tabular}
\end{center}
\end{table*}

A list of key metrics used for graph data benchmarking is shown in Table \ref{table:metrics-for-graph-data}. Here Phase of execution corresponds to at which stage does the metric applies in the benchmark. Data loading corresponds to the stage where input data for the intended processing has been loaded to the system. We will investigate on the benchmarks which are based on the above mentioned metrics next.

\section{Benchmarks for Graph Data Processing Systems}
\label{sec:graph-processing-benchmarks}
Graph processing systems provide a combination of hardware and software to process large graphs efficiently \cite{Dayarathna2017}. Graph processing platforms have significant diversity which results in difficulty of choosing the most suitable platform for implementing an application from a specific application domain \cite{Han:2014:ECP:2732977.2732980}\cite{Iosup:2016:LGB:3007263.3007270}. Benchmarks for graph data processing systems has some important characteristics which can be listed as follows.

Graph data processing system's programming model's performance should be easily characterized using the chosen benchmark. For example, an experimental comparison of systems that use Pregel's vertex-centric approach such as Apache GraphLab \cite{7474329}, Mizan \cite{Khayyat:2013:MSD:2465351.2465369}, GPS \cite{Salihoglu:2013:GGP:2484838.2484843}, and Giraph \cite{Ching:2015:OTE:2824032.2824077} has been conducted by \emph{Han et al.} \cite{Han:2014:ECP:2732977.2732980}. The experiments were conducted on equal grounds considering graph and algorithm centered optimizations. Four different algorithms have been tested on upto 128 Amazon EC2 instances. It was found that Giraph and GraphLab performs well due to their system optimizations.

Data generator of the benchmark suite plays very important role. Lets take an example benchmark for graph processing systems, Waterloo Graph Benchmark (WGB) which has an efficient data generator. WGB generates graphs with real life like properties \cite{10.1007/978-3-319-10596-3_6}. WGB includes a variety of mining, traversal, and operational queries. Its data generator is an efficient and simple one which constructs dynamic graphs. WGB includes a simple and efficient graph data generator which constructs time-evolving graphs which follow the power-law distribution.

The benchmark should be applicable in industrial settings considering large scale systems. For example, GraphBIG is a benchmark suite \cite{7832843} inspired by IBM System G. GraphBIG uses sample data sets, data structures, and workloads from 21 real-world application scenarios. It uses a vertex-centric, dynamic data format, that is broadly used in real-world graph systems. Use of GraphBIG have shown significantly different characteristics throughout different computations as well as extremely irregular memory access patterns \cite{7832843}. They also observed that graph workloads consistently demonstrate large amounts of data sensitivity.

The design of the benchmark should be done considering the best practises established by the community. Beamer \emph{et al.} presented the Graph
Algorithm Platform (GAP) Benchmark Suite \cite{DBLP:journals/corr/BeamerAP15}. The GAP benchmark describes input graphs, graph kernels, and measurement approaches. GAP does not concentrate on loading or building graphs. They have studied the best practices of the community.  They developed the benchmark through graph workload characterization. They mentioned that it is important to generate diverse workload having multiple input graphs and multiple kernels.

\begin{figure}[htbp]
	\centering
		\includegraphics[width=0.7\textwidth]{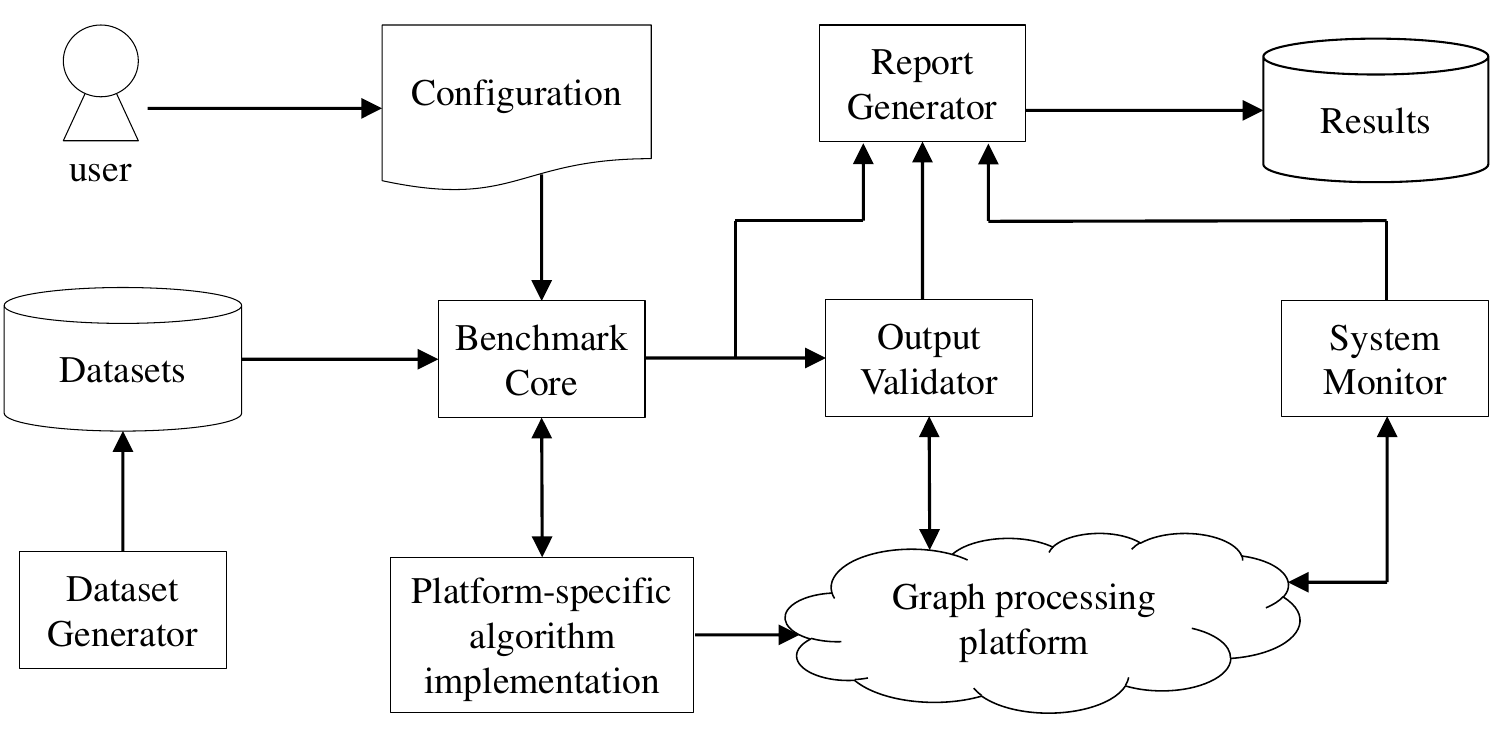}
	\caption{Architecture of Graphalytics benchmark.}
	\label{fig:graphalytics}
\end{figure}

The benchmark should stress various aspects of the benchmarked system. An example for such graph processing platform benchmarks is Graphalytic \cite{Iosup:2016:LGB:3007263.3007270} which comes with an industrial-grade graph analysis benchmark specification. It is intended to match the diversity of operations and the algorithms that are found in graph processing platforms which run distributed manner. The benchmark is designed focusing on 3 vital steps: graph data producer whose output emulates the realistic graph data, algorithm selection via choke point analysis, a ``benchmarking harness'' which consists of an API which allows graph-processing platform developers to attach the benchmark with their platforms easily. The architecture of the Graphalytics benchmark is shown in Figure \ref{fig:graphalytics}. Graphalytics has 6 fundamental algorithms: PageRank, breadth-first search (BFS), weakly connected components (WCC), community detection using label propagation, single source shortest paths (SSSP), and local clustering coefficient. 

The benchmark should also support emerging workloads such as graphs appearing as streams of data. GraphTides is an example for such benchmarks \cite{Erb:2018:GFE:3210259.3210262}. It consists of an approach and an architecture to conduct performance evaluations and measuring the system behavior. The GraphTides framework comprises of a graph stream generator to create graph streams, a testing component, and a results collection unit.

Summarized comparison of benchmarks for graph data processing systems is shown in Table \ref{table:feature-comparison-graph-processing-system-bench}. We indicate the presence of a feature by using a black bullet (\tikz\draw[black,fill=black] (0,0) circle (0.8ex);).  We mark the missing features via a dash (-). When sufficient information could not be found we keep the cell in empty status. We follow this notation across all of the rest of the tables in this paper. Metrics column lists what types of metrics are used for performance measurement by the benchmark experiments. All the benchmarks except GraphTides were batch processing oriented. GraphTides appearing as the latest benchmark framework, as the first one for graph stream processing indicates the rising importance of graph stream benchmarking. Furthermore, most of the benchmarks described in this section used traversal operation. In the next section we will investigate on the details of graph database benchmarks.

\begin{table*}[htbp]
\begin{center}
	  		\caption{Feature comparison of Graph Data Processing System benchmarks.}
  		\label{table:feature-comparison-graph-processing-system-bench}
\begin{tabular}{ | p{0.7cm} | p{2.5cm} | p{1.2cm} | p{1.2cm} | p{8.5cm} |}
\hline
Year & Benchmark & Metrics & Use traversal operation? & Graph algorithms implemented? \\ \hline
2014 & Han \emph{et al.} \cite{Han:2014:ECP:2732977.2732980} & L, M, N & \centering \tikz\draw[black,fill=black] (0,0) circle (0.8ex); & PageRank, SSSP, WCC, Distributed minimum spanning tree (DMST) \\ \hline
2014 & WGB \cite{10.1007/978-3-319-10596-3_6} & L & \centering \tikz\draw[black,fill=black] (0,0) circle (0.8ex); & PageRank, Clustering \\ \hline
2015 & GraphBIG \cite{7832843} & M, CPU & \centering \tikz\draw[black,fill=black] (0,0) circle (0.8ex); & Connected component (CC), Degree centrality, Gibbs inference, Shortest path, Graph coloring, K-core decomposition, Triangle count, Betweenness centrality \\ \hline
2015 & GAP \cite{DBLP:journals/corr/BeamerAP15} & \centering - & \centering \tikz\draw[black,fill=black] (0,0) circle (0.8ex); & PageRank, BFS, CC, SSSP, Triangle Counting (TC), Betweenness Centrality (BC) \\ \hline
2016 & Graphalytics \cite{Iosup:2016:LGB:3007263.3007270} & L, T  & \centering \tikz\draw[black,fill=black] (0,0) circle (0.8ex); & BFS, PageRank, WCC, Local clustering coefficient, Community detection using label propagation, and SSSP \\ \hline
2018 & GraphTides \cite{Erb:2018:GFE:3210259.3210262} & T, CPU & \centering - & - \\ \hline
\end{tabular}
\end{center}
\end{table*}

\section{Benchmarks for Graph Database Systems}
\label{sec:graph-database-benchmarks}
Graph databases (also known as Graph Database Management Systems) are representation, storage, and querying systems for naturally occurring graph structures \cite{6495942}. Specifically a graph database is a storage engine which supports a graph model backed by native graph persistence, with graph primitives and querying techniques \cite{graphdatabsesbook:2013a}. A special feature of a graph database is its schema-free data model. Graph databases are generally optimized for transactional performance and developed with transactional integrity and operational availability targets. Designing graph database benchmarks need to consider multiple performance aspects which affect the operational characteristics of graph databases. These include data size, scalability, highly-availability and reliability, etc \cite{6495942}. The scalability issues happen due to the large data size as well as due to very complex heterogeneous data sources.

Benchmarks developed for graph databases are the pioneers of graph data benchmarking. One such earliest efforts is HPC Scalable Graph Analysis Benchmark (HPCSGAB). It comprises of an application with more than one analysis techniques which uses only one data structure having a directed graph. HPCSGAB has 4 different operations (Graph extraction with BFS, Classification of large vertex sets, Graph construction, and Graph analysis with BC) using power-law graphs \cite{bader:2009a}.

Graph databases have to conduct traversal over the graphs stored on them which is one of their unique features. Ciglan \emph{et al.} \cite{6313678} developed a graph database benchmark focusing on this aspect. The benchmark investigates on conducting traversal operation in an environment having limited amount of memory. The traversal operation has random memory access pattern hence the usual practice has been loading the data set completely into memory. However, this is a significant challenge because the user wants the ability of fast traversal and data persistence simultaneously.

High performance of benchmark system is important aspect which was missing in most of the benchmarks discussed till now. XGDBench is a framework for evaluating performance of graph databases which can run in a distributed compute cluster \cite{6427516}\cite{Dayarathna2014}. XGDBench has been designed targeting social networking service scenarios.  Unlike contemporary graph database benchmarks, XGDBench was designed and implemented as a distributed benchmarking platform which can be scaled in a distributed compute cluster. Furthermore, XGDBench was extended to synthesize workload spikes leveraging its distributed architecture \cite{6906853}. The workload spikes had two categories called volume spikes and data spikes.

Significant growth of systems for storing social network data triggered the need for benchmarks inspired by social networking scenarios. LinkBench is a graph database benchmark published by Facebook \cite{Armstrong:2013:LDB:2463676.2465296}. LinkBench studies the potential of the use of databases as the persistent storage of Facebook. The LinkBench consists of a generator that generates a graph that has similar structural characteristics as Facebook's social graph. 

The standardization efforts in graph processing area resulted in development of benchmarking frameworks which better support multiple graph database servers. Jouili \emph{et al.} developed a distributed graph database comparison framework called GDB. They compared performance of Neo4j, OrientDB, Titan, and DEX with their framework \cite{6693403}. The benchmarking framework is developed targeting Blueprints compliant graph databases and graph servers. GDB consists of three workloads: Load, Traversal, and Intensive workload. The workload simply started a graph database server and loaded it with a particular data set. The traversal workload performs one of the two traversals shortest path, and neighborhood exploration workload. Intensive workload executes a certain number of parallel clients in a synchronous manner. This is very similar to the workload spike generation feature of XGDBench \cite{6906853}. A performance comparison experiment was conducted using Titan, OrientDB, and Neo4j. 

Graph query processing performance becomes very important aspect when it comes to performance of graph databases. Pobiedina \emph{et al.} implemented graph pattern matching on query languages. They had a mix of database systems in three domains SQL, RDF, and graph database \cite{10.1007/978-3-319-10073-9_18}. They mentioned that database systems need to improve their graph pattern matching performance.

\begin{table*}[htbp]
\begin{center}
	  		\caption{Feature comparison of Graph Database benchmarks.}
  		\label{table:feature-comparison-graphdb}
\begin{tabular}{ | p{0.6cm} | p{4cm} | p{1.2cm} | p{1.2cm} | p{1.6cm} | p{5.2cm} |}
\hline
Year & Graph Database Benchmark & Metrics & Property graphs? & Main Programming Language & Graph Generator \\ \hline
2009 & HPC Scalable Graph Analysis Benchmark \cite{bader:2009a} & T  & \centering - &  & R-MAT \\ \hline
2012 & Ciglan \emph{et al.} \cite{6313678} & L & \centering \tikz\draw[black,fill=black] (0,0) circle (0.8ex); & Java & LFR-Benchmark generator \\ \hline
2012 & XGDBench \cite{6427516}\cite{Dayarathna2014} & T & \centering \tikz\draw[black,fill=black] (0,0) circle (0.8ex); & X10 & MAG \\ \hline
2013 & LinkBench \cite{Armstrong:2013:LDB:2463676.2465296} & L, T, CPU, M & \centering  - & Java & Synthetic social graph generator \\ \hline
2013 & GDB \cite{6693403} & L & \centering \tikz\draw[black,fill=black] (0,0) circle (0.8ex); & Java & Barabasi-Albert model \\ \hline
2014 & Pobiedina \emph{et al.} \cite{10.1007/978-3-319-10073-9_18} & L & \centering \tikz\draw[black,fill=black] (0,0) circle (0.8ex); & Python & Synthetic (small-world and erdos renyi networks) and real data \\ \hline
2015 & Beis \emph{et al.} \cite{Beis2015} & L & \centering \tikz\draw[black,fill=black] (0,0) circle (0.8ex); & Java & LFR-Benchmark generator \\ \hline
\end{tabular}
\end{center}
\end{table*}

There are some graph database system benchmarking efforts conducted focusing on specific graph processing scenarios. Beis \emph{et al.} investigated on community detection. They implemented a clustering workload which comprised of Louvain method which is a well-known community detection algorithm \cite{Beis2015}.

A summary of the graph database benchmark comparison is shown in Table \ref{table:feature-comparison-graphdb}. None of the benchmarks had any streaming workloads. The column property graphs indicate whether the graph supports vertex/edge properties. Graph generator is the underlying graph generator used for the benchmark platform. Furthermore, majority of the benchmarks were developed with Java. We list metrics used for evaluation only in Tables \ref{table:feature-comparison-graph-processing-system-bench} and \ref{table:feature-comparison-graphdb} because the metrics specifically used for evaluating graph processing workloads can be easily identified from those works. Next, we explore bigdata benchmarks having graph data processing workloads.

\section{Bigdata Benchmarks with Graph Data Processing Workloads}
\label{sec:bigdata-benchmarks-with-graph-data-processing-capabilities}
Big data processing systems handle the issue of persisting, capturing, analyzing, visualizing, and management aspects of big data. In recent times, several benchmarking platforms came into existence with this purpose. Some of these benchmarks include extensions/workloads for benchmarking graph data processing systems. Some of them have been specifically intended for benchmarking social media /networks.

\subsection{General Bigdata Benchmarks}
\label{subsec:general-bigdata-benchmarks}
It has become very important for general bigdata benchmarks to implement graph data processing workloads as part of their workload suite. BigDataBench is a big data benchmark suite which addresses broad application scenarios, as well as includes variety of different data sets \cite{6835958}. They have used different graph kernels such as BFS and PageRank in their benchmark. They used six representative real-world data sets which include Google Web Graph (directed graph) and Facebook Social Graph (undirected graph). BigDataBench consists of 19 big data benchmark workloads from different dimensions. The workloads include Breadth-first search (BFS), PageRank, Olio Server \cite{6835958}, Kmeans, and Connected Component (CC). Similar to SPARKBENCH \cite{Li:2015:SCB:2742854.2747283},  BigDataBench used Google Web Graph dataset.

BigOP is a benchmarking framework which allows for running comprehensive performance testing \cite{10.1007/978-3-319-05813-9_32}. It automatically generates experiment scenarios having detailed workload patterns for  big data systems. It is also part of the BigDataBench. BigOP consists of a summarized collection of patterns and functions for big data processing. They have implemented PageRank computation as a test example. During the workload generation for PageRank the operations \emph{get}, \emph{filter}, and \emph{transform} are executed iteratively on the data set until a given condition is satisfied. With the PageRank computation implementation they have generated a 3.7 million links and 0.5 million pages which resulted in a 250GB data set. They have found that iterative computation is supported by distributed in-memory computation. They found network communications and frequent disk access are expensive and also further optimizations can be done for relational databases with reference to iterative computation.

\subsection{Social Media/Network Benchmarks}
\label{subsec:social-media-benchmarks}
Systems intended for handling social media needs the capability of storing and processing social networks. Hence the benchmarks targeted for these systems need to also use some kind of graph data storage and processing workloads. BG is a social networking benchmark \cite{10.1007/978-3-319-10596-3_2}. Its intent has been evaluating high performance stores which provides fast responses. BGClient implements a decentralized partitioning technique. It declusters an example data set to N disconnected partitions. However, BG's mix of actions includes only simple social network operations and it does not include traversal operation.

Benchmark for Social Media Analytics (BSMA) is targeted for benchmarking data management systems which conducts analytical queries over social media \cite{Xia:2014:BBA:2733004.2733033}. BSMA provides artificial data generators as well as real-world data sets. Furthermore, it provides a set of workloads. The data generator produces the updates. The workload generator creates queries using query templates. It has significant number of queries including graph temporal queries, operations, aggregate, and hotspot queries. BSMA also consists of a toolkit which allows evaluating and plotting system performance which implements the benchmark.

SPARKBENCH is another example big data benchmark which includes several basic graph analysis workloads \cite{Li:2015:SCB:2742854.2747283}. Graph computation workloads included with SPARKBENCH include three graph algorithms: PageRank, SVD++, and Triangle count. It used Google Web Graph \cite{DBLP:journals/corr/abs-0810-1355} and Amazon Movie Review graphs \cite{McAuley:2013:ACM:2488388.2488466} as the input data sets. Through the workload characterization study they have found that memory is the bottlenecked resource for PageRank and Triangle count algorithms.

The Linked Data Benchmark Council (LDBC) has introduced  Social Network Benchmark (SNB) which models a social network similar to Facebook \cite{10.1145/2723372.2742786}. The dataset comprises of persons and a friendship network that connects them. Based on this dataset three different workloads has been created called SNB-Interactive, SNB-BI, and SNB-Algorithms.

Multi-model databases (MMDB) is an emerging paradigm in database systems which provides a unified platform to manage data stored in different models such as documents, graph, relational, and key-value \cite{Zhang2019ab}. UniBench is a generic benchmark for evaluation of state-of-the-art MMDBs. One of the key challenges they address is generating synthetic multi-model data in such a way it provides correlation in diverse data models. UniBench's workload generator simulates a social commerce scenario which combines social network with the E-commerce context. However, they have modified the LDBC data generator to generate simplified graph data since UniBench's focus is generating multi-model data. Hence, UniBench is not fully fledged graph benchmarking framework.

\begin{table*}[htbp]
\begin{center}
	  		\caption{Feature comparison of BigData benchmarks with Graph Data Processing Workloads.}
  		\label{table:bigdata-bench-with-feature-comparison}
\begin{tabular}{ | p{0.7cm} | p{3cm} | p{1.0cm} | p{1.2cm} | p{1.5cm} | p{6.4cm} |}
\hline
Year & BigData Benchmark & Social Media? & Scalable? & Main Programming Language & Main Use Case \\ \hline
2014 & BigDataBench \cite{6835958} & \centering - & \centering \tikz\draw[black,fill=black] (0,0) circle (0.8ex); & Java & Search engines, social networks, e-commerce \\ \hline
2014 & BigOP \cite{10.1007/978-3-319-05813-9_32} & \centering - & \centering \tikz\draw[black,fill=black] (0,0) circle (0.8ex); &  & Fast storage, Log monitoring, PageRank computation  \\ \hline
2014 & BG \cite{10.1007/978-3-319-10596-3_2} & \centering \tikz\draw[black,fill=black] (0,0) circle (0.8ex); & \centering \tikz\draw[black,fill=black] (0,0) circle (0.8ex); & Java & Social networks \\ \hline
2014 & BSMA \cite{Xia:2014:BBA:2733004.2733033} & \centering \tikz\draw[black,fill=black] (0,0) circle (0.8ex); & \centering \tikz\draw[black,fill=black] (0,0) circle (0.8ex); & Java & Social network  \\ \hline
2015 & LDBC SNB \cite{10.1145/2723372.2742786} & \centering \tikz\draw[black,fill=black] (0,0) circle (0.8ex); & \centering \tikz\draw[black,fill=black] (0,0) circle (0.8ex); & Java & Social Network \\ \hline
2015 & SPARKBENCH \cite{Li:2015:SCB:2742854.2747283} & \centering - & \centering \tikz\draw[black,fill=black] (0,0) circle (0.8ex); & Scala & Machine Learning and Graph computation \\ \hline
2019 & UniBench \cite{Zhang2019ab} & \centering \tikz\draw[black,fill=black] (0,0) circle (0.8ex); & \centering \tikz\draw[black,fill=black] (0,0) circle (0.8ex); & Java & Social networks, e-commerce \\ \hline
\end{tabular}
\end{center}
\end{table*}

Summary of the feature comparison between the bigdata benchmarks with graph data processing workloads is shown in Table \ref{table:bigdata-bench-with-feature-comparison}. Unlike the benchmarks specifically intended for graph data processing and management all of the bigdata benchmarks having graph data processing workloads were scalable. This is an expected behavior from any bigdata benchmark. Workload generators are the fundamental components of the benchmarks which generate the required workloads for benchmarking process which are described next.

\section{Workload Generators}
\label{sec:workload-generation}
Graph data processing benchmarks are dependent on workload generators for constructing scalable workloads. Workload generators represent the Data Generation phase of graph data processing system benchmarking process. The workload generator algorithms should emulate meaningful real world processing \cite{capotagraphalytics}. There are two main aspects of workload generation. First, realistic synthetic graph data need to be generated. Second, the generated data need to be sent to the target in a systematic way (i.e., workload emission). Next, we present notable synthetic graph generators.

\subsection{Synthetic Graph Generators}
\label{subsec:graph-generation}
Properties of a realistic synthetic data set include small diameter, power law degree distribution, high clustering coefficient \cite{6313678}. The graph data sets used should represent real-world graphs \cite{Dayarathna2014}\cite{6906853} while the graphs should be suitable of processing on different scales \cite{capotagraphalytics}.

Among the graph data generators, Stochastic Kronecker Graph (SKG) and R-MAT models have attracted significant interest of the benchmarking community due to their simplicity and their ability to represent many properties of real-world graphs \cite{DBLP:journals/corr/abs-1208-2239}. For example, Graph 500 which is one of the famous benchmarks for graph data generates SKGs of different scales and evaluates the time to conduct breadth-first search \cite{graph500:2010a}\cite{6114175}.

There have been considerable interest on reflecting the real world graph features on artificial graph data. If we take social networks, individuals who are connected through friendship relationships exhibit ``relational correlation'' \cite{pfeiffer2014attributed}. Multiplicative Attribute Graphs (MAG) is one example of data generator which ensures the realistic community structure of synthetics graphs \cite{10.1007/978-3-642-18009-5_7}. The LFR Benchmark and S3G2 benchmarks are some other examples for such benchmarks \cite{PhysRevE.78.046110}\cite{10.1007/978-3-642-36727-4_11}.

The generated data set needs to emulate real world data handled by the target applications. Ming \emph{et al.} investigated about creating graphs to achieve the 4V requirements of big data \cite{10.1007/978-3-319-10596-3_11}. They created a Big Data Generator Suite (BDGS) to generate synthetic data while protecting their inherent characteristics. BDGS is part of the BigDataBench project and it focuses on nineteen workload scenarios as well as the 6 applications from its mother project.

Obtaining a graph data generator that accurately matches with the requirements of a given benchmark is quite challenging task. DataSynth allows for generating property graphs with unique characteristics via the use of customizable schemas \cite{Prat-Perez:2017:TPG:3078447.3078453}. They have used a property-to-node matching algorithm. DataSynth is also capable of regenerating user-given property-structure correlations and distributions of property values which results in property graphs having similar structure to real world graphs.

\begin{table*}[htbp]
\begin{center}
	  		\caption{Feature comparison of graph generators.}
  		\label{table:graph-generators-feature-comparison}
\begin{tabular}{ | p{0.7cm} | p{5cm} | p{1.2cm} | p{4cm} | p{1.4cm} | p{1.2cm} |}
\hline
Year & Workload Generator which incorporates the graph data generator & Scalable? & Data Generator Type & Main Programming Language & Support for Rich Graphs \\ \hline
2010 & Graph500 Generator \cite{graph500:2010a} & \centering \tikz\draw[black,fill=black] (0,0) circle (0.8ex); & SKG & C/C++ & - \\ \hline
2012 & XGDBench Incremental Workload Generator \cite{6906853} & \centering \tikz\draw[black,fill=black] (0,0) circle (0.8ex); & MAG & X10 & \tikz\draw[black,fill=black] (0,0) circle (0.8ex); \\ \hline
2014 & BDGS \cite{10.1007/978-3-319-10596-3_11} & \centering \tikz\draw[black,fill=black] (0,0) circle (0.8ex); & Kronecker & C/C++ & \tikz\draw[black,fill=black] (0,0) circle (0.8ex); \\ \hline
2017 & DataSynth \cite{Prat-Perez:2017:TPG:3078447.3078453} & \centering \tikz\draw[black,fill=black] (0,0) circle (0.8ex); & Property generator and Structure generator &  & \tikz\draw[black,fill=black] (0,0) circle (0.8ex); \\ \hline
2017 & TrillionG \cite{Park:2017:TTS:3035918.3064014} & \centering \tikz\draw[black,fill=black] (0,0) circle (0.8ex); & recursive vector model & Scala/Java & \tikz\draw[black,fill=black] (0,0) circle (0.8ex); \\ \hline
\end{tabular}
\end{center}
\end{table*} 

TrillionG is a massive graphs generator which constructs large graphs quickly utilizing only few memory \cite{Park:2017:TTS:3035918.3064014}. Their proposed Recursive Vector graph generation model. Orders of magnitude performance improvements have been observed with TrillionG with respect to the contemporary data generators. TrillonG generates massive graphs in short duration and it uses less memory. A graph with one trillion edges can be generated using R-MAT and Kronecker models in 2 hours with ten personal computers.

Summary of the comparison between the graph data generators is shown in Table \ref{table:graph-generators-feature-comparison}. All of the graph generators surveyed were scalable and most of them supported rich graphs.

\subsection{Workload Emission}
\label{subsec:workload-emission}
Once the synthetic data set has been generated, the next step is to inject the workload to the target system. The workload emission pattern determines the behavior of the benchmarked system during the benchmark's execution time. The workload should also exercise the ``choke points'' \cite{capotagraphalytics} of the system being tested. Choke-point is an important technological challenge identified by domain experts such as system architects in relation to the benchmarked system's performance. The choke-points identified for Graphalytics include: Excessive network utilization, Large graph memory footprint, Poor access locality, Skewed execution intensity.

Certain workload emitters create step-wise incremental workloads. XGDBench's step-wise incremental workload generator is an example system for this category \cite{Dayarathna2017}\cite{6906853}. The workload can be specifically targeted at some data points. If not the entire graph can be targeted. Next, we describe workload characterizations conducted using the graph benchmarks.

Recently, there have been work done on graph generation which takes into account the complete query workloads on which the generated graphs are intended to be executed against. This unified approach leverages the fact that the problems of graph and query workload generation are interactable with each other. gMark is an example framework of this category which follows a schema-driven approach for generating synthetic graph instances linked with sophisticated query workloads \cite{7762945}. gMark is known to be the first benchmark which generates workloads involving recursive path queries \cite{7762945}.

\section{Workload Characterization on Graph Data Processing Systems}
\label{sec:graph-data-processing-workload-characterization}
Comparison of multiple graph data processing system performance is an important activity for understanding the behavior and the important system requirements of graph data systems. This process is known as workload characterization and it corresponds to the execution stage of the graph data processing system benchmarking process. 

We separate work done on workload characterization of graph data processing systems into three areas called workload characterizations on Non-GPU systems, workload characterizations on GPU systems, and workload characterizations on many core processors.

\subsection{Workload Characterization on Non-GPU based Graph Data Processing Systems}
\label{subsec:graph-data-processing-workload-characterization-non-gpu}
Non-GPU systems are computer systems which do not include GPUs. These systems only consist of/use CPUs to provide the processing capabilities. Majority of graph data processing systems are commodity server based systems and they are non-GPU systems.

\subsubsection{Workload Characterizations Focusing on Programming Models}
\label{subsubsec:graph-data-processing-workload-characterization-non-gpu-programmingmodel}
Importance of conducting workload characterizations on large scale real-world non-GPU systems has been emphasized in multiple work. Redekopp \emph{et al.} presented analysis and optimizations for complicated graph analysis algorithms \cite{6569812}. They evaluated graph processing frameworks which follow Bulk Synchronous Parallel Processing (BSP) model on cloud. They analyzed the impact of graph partitioning for BSP frameworks. Their findings indicate that although inter-worker communication can be reduced with the use of partitioning, this may result with imbalanced workload which reduces the effectiveness of the use of partitioners by BSP frameworks. They evaluated the effectiveness of graph partitioning with METIS partitioner, a heuristic streaming partitioner, and simple hashing of vertices. They have run PageRank, Betweenness-centrality (BC), and All Pairs Shortest Path (APSP) over a Google web graph as well as over a citations-Patents graph.

In most of the graph processing systems developer productivity has been kept as the second priority compared to system performance. For example, Satish \emph{et al.} compared graph analytics frameworks using massive graph data sets. They have used GraphLab \cite{7474329}, CombBLAS, SociaLite, Galois, and Giraph \cite{Ching:2015:OTE:2824032.2824077} in their study \cite{Satish:2014:NMG:2588555.2610518}. The purpose of their work was to develop a road map to enhance the performance of different graph frameworks so that they could alleviate their performance gaps. They mentioned that once the performance differences have been mitigated the selection of the framework then depends on the productivity issues.

Significant focus has been made on characterizing performance of Pregel graph processing model. Han \emph{et al.} conducted a performance study on Pregel-like systems. Performance reported by multiple graph processing systems is not easily comparable. They address this issue by working on to expose the behavior of different systems that are attributable to both built-in and optional system optimizations which are intrinsic to the system and agnostic to the graph or algorithm used \cite{Han:2014:ECP:2732977.2732980}. They used 4 algorithms: SSSP, distributed minimum spanning tree, PageRank, and weakly connected components for their study. They compared GPS \cite{Salihoglu:2013:GGP:2484838.2484843}, Giraph, GraphLab, and Mizan \cite{Khayyat:2013:MSD:2465351.2465369} on similar basis by evaluating algorithm and graph neutral optimizations with multiple metrics. They found GraphLab's and Giraph's synchronous mode give satisfactory overall performance. GPS is better with use of memory in an efficient manner.

Workload characterizations have proven the fact that different systems show different performances even if they follow the same graph processing model. Lu \emph{et al.} conducted multiple evaluations to investigate graph processor performance having variety of features and on algorithms with various logical designs \cite{Lu:2014:LDG:2735508.2735517}. They studied about vertex centric graph processors such as Giraph \cite{Ching:2015:OTE:2824032.2824077}, GraphLab (i.e., PowerGraph) \cite{7474329}, GPS \cite{Salihoglu:2013:GGP:2484838.2484843}, Pregel+ \cite{Yan:2015:ETM:2736277.2741096}, and GraphChi \cite{Kyrola:2012:GLG:2387880.2387884}. They also found that Pregel+ \cite{Yan:2015:ETM:2736277.2741096} and GPS \cite{Salihoglu:2013:GGP:2484838.2484843} have better overall performance compared to Giraph and GraphLab. They also found specific reasons for why certain graph processing systems have better performance compared to the others in different cases. For example, due to the use of techniques such as mirroring, message combining, and request-respond techniques Pregel+ has better performance.

There are workload characterizations conducted on much wider scopes spanning beyond single programming model. Guo \emph{et al.} described practical technique of evaluating characteristics of graph processors with the aim of addressing the question ``How well do graph processing platforms perform?'' \cite{Guo:2014:WGP:2650283.2650530}. Their method depends on the ability of creating a detailed process for evaluation. It also depend on identifying representative algorithms, datasets, and metrics for verifying the important features of such platforms. The important aspects they mentioned are resource utilization, execution time, overhead, and vertical and horizontal scalability. Their work almost resembles a benchmark suite since they selected and implemented five algorithms (general statistics, graph evolution, BFS, Community detection, and CC) with 7 datasets from different application areas. Furthermore, they implemented this benchmark suite on 6 famous platforms: YARN, Hadoop, Giraph, Stratosphere, Neo4j, and GraphLab.

All the above mentioned studies have used a variety of graph data sets and algorithms. However, none of the work tried to find what is the minimal set of graph algorithms and data sets which could deliver the required quality of benchmarking experience. Yang \emph{et al.} tried to address this issue \cite{Yang:2015:UGC:2749246.2749257}. In order to identify the robustness of the performance studies Yang \emph{et al.} characterized the characteristics change across graph structures and algorithms at various sizes. They found that these characteristics form a very broad space with up to 1000-fold variation. They also mentioned that any inefficient study of this space will result in ad-hoc studies and limited understanding. Therefore, they constructed an ensemble of graph analysis tasks and defined 2 metrics to evaluate how completely and efficiently an ensemble evaluates the space. They found that evaluations restricted to a single graph or to a single algorithm will unevenly evaluate a graph processing system. They also found that evaluating graph and algorithm diversity may considerably enhance the quality (200\% more efficient and 30\% more complete). They mentioned that they can reduce the complexity the user has to under go considering factors such as graphs, algorithms, and runtime while simultaneously preserving the quality of benchmarking.

A recent work on Parallel algorithms emphasizes the use of graph primitives for mapping, filtering, reducing and packing the neighbours of a vertex \cite{}.

\subsubsection{Workload Characterizations Focusing on Partitioning Strategies}
\label{subsubsec:graph-data-processing-workload-characterization-non-gpu-partition}

Experimental comparisons of the partitioning techniques present in three distributed graph processing systems (PowerGraph, GraphX, and PowerLyra) was made by Verma \emph{et al.} \cite{Verma:2017:ECP:3055540.3055543}. Through the performance experiments they found that no single partitioning strategy is the best fit for all situations. They also found that the selection of partitioning technique has a considerable effect on resource usage and application run-time. They found that for PowerGraph the replication factor is a good indicator of partitioning quality. This is because it is linearly correlated with network usage, computation time, and memory utilization. They also presented two decision trees to help users of the systems to pick a partitioning strategy.

In another similar work Guo \emph{et al.} modeled the time complexity of graph processors considering partitioning techniques \cite{GUO2017106}. They also identified the deficiency of performance studies considering partitioning policies. They mentioned that narrowly designed experiments may misreport performance by considerable amount. This could especially happen when the algorithms and the input workloads change from the situations evaluated in some studies. They have used multiple metrics to evaluate twelve partitioning policies. The experiments were conducted using 3 big graph data sets on 3 platforms. Their results indicated that their novel database partitioner approach indicates better results. However, the contemporary stream-based graph partitioning policies such as Combined Policy (CB) and Linear-weighted Deterministic Greedy Policy (LDG) do not perform well.

\begin{table*}[htbp]
\centering
\begin{center}
	  	\caption{Comparison of graph workload characterizations on Non-GPU systems.}
  		\label{table:graph-workload-characterization}
\begin{tabular}{ | p{0.6cm} | p{2.8cm} | p{6.4cm} | p{5cm} |}
\hline
Year & Investigator(s) & Scenario/Graph Algorithms & Data sets \\ \hline
2013 & Redekopp \emph{et al.} \cite{6569812} & BC, PageRank, and APSP & LiveJournal (LJ) social networks, SlashDot (SD), patent citation network (CP), and a web graph from Google (WG) \\ \hline
2014 & Satish \emph{et al.} \cite{Satish:2014:NMG:2588555.2610518} & PageRank, BFS, Triangle Counting, Collaborative Filtering & Facebook, Wikipedia, LiveJournal, Netflix, Twitter, Yahoo Music, Synthetic, Graph500, Synthetic Collaborative Filtering \\ \hline
2014 & Han \emph{et al.} \cite{Han:2014:ECP:2732977.2732980} & Single-source shortest path (SSSP), Graph analysis workloads such as PageRank, Distributed minimum spanning tree (DMST), and weakly connected components (WCC) & soc-LiveJournal1 (LJ), com-Orkut (OR), arabic-2005 (AR), twitter-2010 (TW), uk-2007-05 (UK); \\ \hline
2014 & Lu \emph{et al.} \cite{Lu:2014:LDG:2735508.2735517} & Diameter Estimation, PageRank, Single-Source Shortest Paths (SSSP), HashMin, Shiloach-Vishkin’s Algorithm (SV), Bipartite Maximal Matching (BMM), Graph Coloring (GC) & Friendster, Twitter, LiveJournal, BTC, USA Road \\ \hline
2014 & Guo \emph{et al.} \cite{Guo:2014:WGP:2650283.2650530} & General statistics, BFS, Connected Component, Community detection, graph evolution & Amazon, WikiTalk, KGS, Citation, DotaLeague, Synth, Friendster \\ \hline
2015 & Yang \emph{et al.} \cite{Yang:2015:UGC:2749246.2749257} & Connected Components (CC), Triangle Counting (TC), KC, PageRank (PR), Approximate Diameter (AD), and Single-Source Shortest Path (SSSP) & Scale-free synthetic networks \\ \hline
2016 & Wei \emph{et al.} \cite{7474329} & PageRank, WCC, and BFS & soc-LiveJournal1 and Twitter rv \\ \hline
2017 & Verma \emph{et al.} \cite{Verma:2017:ECP:3055540.3055543} & PageRank, WCC, K-Core Decomposition, SSSP, Simple Coloring & road-net-CA, road-net-USA, LiveJournal, Enwiki-2013, Twitter, UK-web \\ \hline
2017 & Zhang \emph{et al.} \cite{Zhang:2017:AIP:3127479.3128606} & SSSP, HashMin, PageRank, Triangle Counting & Twitter, Youtube, Orkut, Friendster and road network \\ \hline
2017 & Guo \emph{et al.} \cite{GUO2017106} & BFS, PageRank, and WCC & Twitter, Scale 26, Datagen\_p10m \\ \hline
2017 & Shantharam \emph{et al.} \cite{7965151} & BFS, PageRank, Connected Components, K-Core & RMAT-27, Twitter, SK2005, Wiki \\ \hline
2018 & Ammar \emph{et al.} \cite{Ammar:2018:EAD:3231751.3242935} & PageRank, WCC, SSSP, K-hop & Twitter, WRN, UK200705, ClueWeb \\ \hline
\end{tabular}
\end{center}
\end{table*}

\subsubsection{Workload Characterizations Focusing on Cloud and Novel Storage Techniques}
\label{subsubsec:graph-data-processing-workload-characterization-non-gpu-cloud}
Workload characterization studies also explored the performance behaviors of cloud based graph processing. Wei \emph{et al.} conducted a comparison of Apache Spark and GraphLab \cite{7474329}. The benchmark suite they used included example data sets and algorithms to compare the computing engines' performance. They tested the benchmark suite on VMs on cloud in addition to testing them locally. Unlike the previous studies they also did experiments to measure the possible performance reduction due to the availability of one processing engine. They observed GraphLab performed better in graph processing use cases compared to Spark while it had similar performance behavior to Spark for non-graph algorithms. Running graph algorithms on cloud VMs using Spark resulted in doubled elapsed time compared to running the same setup on local compute clusters.

More recent workload characterizations correlated the cost of operating cloud computing systems with graph processing system performance. Zhang \emph{et al.} did detailed study of eight graph analytics systems with varying configurations on centralized vs distributed deployment, and out-of-core vs in-memory deployments \cite{Zhang:2017:AIP:3127479.3128606}. They consider other factors other than performance such as cost in their study. They mentioned that there is no silver bullet like solution. They mentioned that centralized, in-memory systems may perform better with certain workloads since memory is cheaper today. Therefore, if the entire data set can be loaded into the memory, then in-memory system would be good choice although high throughput is offered by distributed graph processing systems. In elastic data processing systems \cite{Ravindra:2017:LAE:3030207.3030227} and in a pay-as-you-go cloud model, initial sunk-cost is less important compared to the latencies. In some situations users would prefer high throughput compared to low latency results which may also remove the requirement for running the system on high-end hardware.

In a completely different study which involved byte-addressable Non-Volatile Memory (NVM) Shantharam \emph{et al.} conducted a quantitative assessment of the performance of large scale-free graph processing. The graphs were stored in NVM \cite{7965151}. Their results indicated that locality aware algorithm processing and data structure layouts can provide better results with NVMs with minimal performance impact. This makes NVM highly valuable for large graph processing.

In another Amazon EC2 cloud based performance study, Ammar \emph{et al.} evaluated HaLoop, Hadoop, Giraph, Vertica, GraphLab (PowerGraph), Flink Gelly, GraphX (SPARK), and Blogel on 4 large data sets \cite{Ammar:2018:EAD:3231751.3242935}. Experiments were conducted on four datasets. Relative performance comparison between these systems have been presented as the conclusions. They also concluded that programming language's impact on the graph processing system performance is yet to be studied. They mentioned that the best system varies according to the workload and input graph data for the system.

Table \ref{table:graph-workload-characterization} presents a summary of the graph workload characterizations on non-gpu systems. Note that the table preserves chronological order while items in the subsections may not. BFS and PageRank have been the choice for multiple graph workload characterizations. Furthermore, almost all of these studies have been conducted following social networking scenarios or using social network data sets which indicates the importance of studying this use case further in relation to system performance. Next, we investigate on workload characterization of GPU based systems for graph data processing.

\subsection{Workload Characterization on GPU based Graph Data Processing Systems}
\label{subsec:graph-data-processing-workload-characterization}
GPU based data processing is relatively new area of graph data processing with increasing interest \cite{6799136}. For example, nvGRAPH is a graph analytics library from NVIDIA which consists of parallel algorithms for conducting analytics on graphs with up to two billion edges \cite{nvgraph:2018a}.

Graph data processing applications are type of irregular applications. This is because these applications use operations which traverse, update, and build data structures such as priority queues, graphs, and trees which are irregular. Furthermore, these applications are irregular because of their data dependent memory-access patterns. Burtscher \emph{et al.} conducted a workload characterization of set of real-world GPU applications which are irregular \cite{Burtscher:2012:QSI:2473499.2474126}. They found that there are other major factors other than irregularity which determines system performance.

\begin{table*}[htbp]
\centering
\begin{center}
	  	\caption{Comparison of graph workload characterizations on GPU systems.}
  		\label{table:graph-workload-characterization-gpu}
\begin{tabular}{ | p{0.7cm} | p{1.8cm} | p{4cm} | p{4.6cm} | p{3cm} |}
\hline
Year & Investigator(s) & Graph Algorithms & Data sets & GPU Model used \\ \hline
2012 & Burtscher \emph{et al.} \cite{Burtscher:2012:QSI:2473499.2474126} & BFS, SSSP & Road network data set ((23 M nodes, 58 M edges) & 1.45 GHz Quadro 6000
GPU having a memory of 6 GB. The GPU has 448 cores \\ \hline
2013 & Che \emph{et al.} \cite{6704684} & Connected components labeling (CCL), Dijkstra (DJK), Graph coloring (CLR), Maximal independent set (MIS), Floyd-Warshall (FW), Friend recommendation (FRD), Betweenness centrality (BC), Page rank (PRK) & GTgraph random-graph generator to generate graphs & AMD Radeon HD 7950 (Tahiti) discrete GPU \\ \hline
2015 & Guo \emph{et al.} \cite{guo:ccgrid2015a} & BFS, PageRank, WCC & Amazon, WikiTalk, Citation, KGS, DotaLeague, Scale22, Scale23, Scale24, Scale25 & Nvidia Tesla K20m GPU (onboard 5 GB memory), Nvidia GeForce GTX 580 GPU (onboard 3 GB memory per GPU), Nvidia GeForce GTX 480 GPU (onboard 1.5 GB memory)  \\ \hline
2017 & Liu \emph{et al.} \cite{DBLP:journals/corr/abs-1708-04701} & Single Source Shortest Path, Breadth First Search, Graph Coloring, kCore, Triangle Count, PageRank & roadNet\_TX, roadNet\_CA, amazon0312, ego-Twitter, delaunay\_n18, delaunay\_n19, hugetrace-00000, hugetric-00000, kron\_g500-logn17, kron\_g500-logn18, rgg\_n\_2\_16\_s0, rgg\_n\_2\_19\_s0 & Nvidia Geforce GTX1070 GPU \\ \hline
\end{tabular}
\end{center}
\end{table*}

Performance studies in this area also uses variety of graph algorithms/domains. Che \emph{et al.} presented and characterized a suite of GPGPU graph applications called \emph{Pannotia} which is implemented in OpenCL \cite{6704684}. The purpose of their study was to provide an overall picture of the range of characteristics, similarities, and differences among a broad set of graph applications so that the system designers could understand the characteristics and performance bottlenecks. Pannotia consists of a variety of applications which present diverse parallelism and access patterns and varying GPU resource usage characteristics. They conducted a characterization of the graph applications on real GPU hardware and used hardware counters available in an AMD GPU for this study. They also did a hierarchical clustering analysis to characterize the range of benchmark behaviors and also studied their sensitivity to diverse graph structures. They observed that different program-input pairs may show completely different characteristics.

An practical technique for studying about GPU-enabled graph processing systems was proposed by Guo \emph{et al.} \cite{guo:ccgrid2015a}. The performance evaluation methodology they proposed for GPU based systems include a selection of new data sets, new performance metrics, and algorithms. They have used 3 typical graph-processing algorithms and 9 different graphs and  on the three GPU-enables systems Medusa, Totem, and MapGraph. Their study involved analysis of GPU-enabled systems.  The results described aspects such as raw processing power, scalability, as well as the performance effects made by the optimizations and the use of GPUs.

Summary of graph workload characterizations on GPU systems is shown in Table \ref{table:graph-workload-characterization-gpu}. Similar to the summary shown in Table \ref{table:graph-workload-characterization}, most of the graph workload characterizations on GPU systems have used BFS and PageRank as their graph algorithms. GPUs are naturally suitable for matrix multiplication operations. However, the copy overhead between the CPU and GPU have been significantly high which makes it difficult to adapt GPUs for graph processing.

\subsection{Workload Characterization on Manycore Processor based Graph Data Processing Systems}
\label{subsec:graph-data-processing-knl}

Manycore processors allow for using standard programming languages and APIs for implementing systems having similar performance for GPU based systems. They share multiple similarities to GPUs but unlike GPUs which are co-processors, they are fully functional processing units.

Performance study using Intel Xeon Phi (2nd generation) processor - KNL on multi-threaded graph applications was done by Liu \emph{et al.} \cite{DBLP:journals/corr/abs-1708-04701}. Specifically they empirically evaluated a cutting edge graph framework on 3 hardware platforms: Nvidia Tesla P40 GPU, KNL MIC processor, and Intel Xeon E5 CPU. They also demonstrate that the KNL displays significant power efficiency and performance with multi-threaded graph applications. Furthermore, they characterized the performance details of KNL on multi-threaded graph benchmarks. They measured and analyzed the effectiveness of the architectural improvements of KNL. They observed that distinctive numbers of threads are required by different graph applications in order to perform best in terms of performance.

Studies have shown the importance of leveraging the hardware features of manycore processors when running graph algorithms on such systems. Uta \emph{et al.} studied the performance behavior between KNL and graph processing platforms \cite{8514860}. They have found that a hardware dimension of complexity is added by the KNL-based infrastructure. They mentioned that if a graph-processing platform uses KNL it needs to use memory located on the chip, enhance communication between machines, and enhance the use wide-vectors.

This section was on workload characterization on graph processing systems. Next, we will investigate on workload characterizations conducted on graph data management systems.

\section{Workload Characterization on Graph Data Management Systems}
\label{sec:graph-data-management-workload-characterization}
Graph databases has undergone significant architecture improvements over the last decade. Similarly graph database workload characterizations have also evolved significantly over the last decade time period. One of the earliest works on this area called the HPC Scalable Graph Analysis Benchmark was done by Dominguez-Sal \emph{et al.} \cite{Dominguez-Sal:2010:SGD:1927585.1927590}. They evaluated 4 graph databases. An important fact to note is that they do not attempt to validate the results of benchmarking complete-scale runs. This is because at full scale the dataset size is huge. Therefore, its exact characteristics relies on the used pseudo-random number generator. Hence the verification process uses soft checking of the results.

Significant discussions have been made between the choice of graph database vs RDBMS technologies for storing graph data. Performance of Neo4j and MySQL was evaluated by Vicknair \emph{et al.} \cite{Vicknair:2010:CGD:1900008.1900067} in the context of storing graphs. They mentioned that relational databases are generally efficient unless the data comprises of relationships  that require joins of large tables. The biggest data set they used was 100k nodes and data was generated using a synthetic data generator. It was found that both Neo4j and MySQL databases performed similarly. However, with structural queries Neo4j was faster.

Similar to graph processing benchmarks, social networking has been used in multiple graph database benchmarks. For example, Dominguez-Sal \emph{et al.} work can be taken as an example \cite{Dominguez-Sal:2010:DDG:1946050.1946053}.
The study was conducted on multiple application types such as proteomics, travel planning and routing, social network analysis, recommendation systems. The study concluded that social network analysis can be regarded as one of the representative applications for general graph database applications.

Unlike graph processing frameworks, graph databases use some kind of query language to specify operations conducted on graphs. Hence performance measurement at the query execution engine is very important. Study of query languages such as Gremlin \cite{website:gremlin:2013a}, Neo4j native API \cite{Holzschuher:2013:PGQ:2457317.2457351}, and Cypher \cite{neo4jbook:2012a} was made by Holzschuher \emph{et al.}. The study was focused around Apache Shindig \cite{website:shindig:2013a}. They used Neo4j instead of the relational back-end of Shindig for querying data.

Open source graph databases' performance comparison was conducted by McColl \emph{et al.}. They tested four key graph algorithms on 12 open source graph databases on graphs up to 256 million edges \cite{McColl:2014:PEO:2567634.2567638}. The well known algorithms they tested include Shiloach Vishkin connected components algorithm (SV), PageRank (PR), and SSSP. They used R-MAT graph generator in their experiments. A notable feature of their study was that they conducted qualitative comparisons of the graph database systems such as the ease of setup and use, developer experience, etc.

\begin{table*}[htbp]
\centering
\begin{center}
	  	\caption{Comparison of graph workload characterizations on graph data management systems.}
  		\label{table:graph-workload-characterization-graph-data-management-systems}
\begin{tabular}{ | p{0.7cm} | p{2.4cm} | p{5.4cm} | p{4cm} | p{1.6cm} |}
\hline
Year & Investigator(s) & Types of workloads involved & Data sets & Distributed? \\ \hline
2010 & Dominguez-Sal \emph{et al.} \cite{Dominguez-Sal:2010:SGD:1927585.1927590} & Data Loading, Scanning the edges, 2-hops traversal, BC & R-MAT scales 10, 15, 20, 22, 24 & - \\ \hline
2010 & Vicknair \emph{et al.} \cite{Vicknair:2010:CGD:1900008.1900067} & Graph traversal (structural query) and count the number of nodes (data query) & Graphs were created with 1000, 5000, 10000, and 100000 nodes & - \\ \hline
2010 & Dominguez-Sal \emph{et al.} \cite{Dominguez-Sal:2010:DDG:1946050.1946053} & Graph Analysis, Traversals, Communities, Graph Anonymization, Components, Pattern matching, Centrality Measures, and Others (e.g., finding similarity between nodes) & Social network, Protein Interaction, Recommendation, Routing & - \\ \hline
2013 & Holzschuher \emph{et al.} \cite{Holzschuher:2013:PGQ:2457317.2457351} & Suggesting friendships, group memberships, lists of friends, activities, social net data, and retrieval of person profiles & Slashdot data set from SNAP & - \\ \hline
2014 & McColl \emph{et al.} \cite{McColl:2014:PEO:2567634.2567638} & PR, CC, SSSP & R-MAT graphs with 32K (small), 1K (tiny), 16M (large), and 1M (medium) vertices with 256K, 8K, 256M, and 8M undirected edges & \tikz\draw[black,fill=black] (0,0) circle (0.8ex); \\ \hline
2014 & Wycislik \emph{et al.} \cite{Wycislik:2014a} & Local Clustering Coefficient (LCC) & Three artificially generated networks and one real life network & - \\ \hline
2017 & TigerGraph \cite{tigerperformance} & k-neighborhood query, Weakly Connected Component, PageRank & graph500-22 (2.4M vertices and 64M edges), twitter\_rv.net (41.6M vertices, 1.47B edges) & - \\ \hline
2018 & Lissandrini \emph{et al.} \cite{10.14778/3297753.3297759} & Data loading, Creation,Read,Update,Deletion (CRUD) operations on nodes, edges, and properties, Traversal & Three real world networks (Yeast, MiCo, and Freebase(Three versions called Frb-(O/S/M/L) were derived)) and one synthetic network & - \\ \hline
\end{tabular}
\end{center}
\end{table*}

Another performance comparison between a graph database and a relational database was conducted by Wycislik \emph{et al.} \cite{Wycislik:2014a}. The motivation behind their workload characterization was to uncover whether social graph storage is more efficient or not in Neo4j or in a relational database such as Oracle. In their study they counted the duration for calculating local clustering coefficient. They observed that Neo4j is faster with real life network. Since the filtering on entity features is a basic function of RDBMS they found that Oracle performed better when Local Clustering Coefficient (LCC) was calculated.

Performance study of TigerGraph, Neo4j, and Titan was made by TigerGraph team in \cite{tigerperformance}. They tested the TigerGraph 0.8 with Neo4j 3.1.3 community edition, and TitanGraph 1.0.0. The tests were carried out in Amazon EC2 cloud. They found that TigerGraph has 4 times to 100 times faster graph traversal and query response times compared to both Neo4j and Titan. They also observed that query speed increases as additional computers are added achieving a 4.7 times speed increase with 8 computers. They used graph500 data generator to generate scale 22 graph as well as the Twitter network dataset from KAIST \cite{Kwak10www}.

A microbenchmarking based approach for graph database benchmarking was presented by Lissandrini \emph{et al.} \cite{10.14778/3297753.3297759}. They derived a set of primitive operators (i.e., primitive queries) decomposing the complex queries typically found in macrobenchmarks. Their work was focused only on single machine installation and all the systems they used for their benchmark evaluation supported Apache Tinker Pop framework. Detailed performance experiments were conducted using seven systems. They emphasized the equal importance of the use of microbenchmarking along with macro (complex/application) benchmarks when evaluating graph database performance.

Table \ref{table:graph-workload-characterization-graph-data-management-systems} shows a summary of the comparison between graph workload characterizations on graph data management systems. Most of these workload characterizations focus only on graph algorithms which needs to be improved. Graph data management systems have to cope with other concerns such as data loading and persistence. Such issues have been discussed by only few workload characterization studies. Next section investigates on standardization efforts made on benchmarking graph data management and processing systems.

\section{Standardization Efforts}
\label{sec:standardization}
Most of the initial efforts on benchmarking graph data systems were isolated efforts. However, recently, there have been initiatives for setting up independent bodies for coordinating the efforts for development of graph benchmarks.

One such standardization efforts with benchmarking data intensive supercomputing applications is called Graph500 \cite{4086145}. Graph 500 has a compact usecase which has various analysis techniques working on a single data structure. There have been multiple research carried out to characterize the performance of Graph500 benchmark and to improve its performance.

Another initiative is the Linked Data Benchmarking Council (LDBC) which develops industry-strength benchmarks targeting RDF and graph data management systems \cite{Angles:2014:LDB:2627692.2627697}. The LDBC project has developed two benchmarks: a social network use case based graph data management benchmark (i.e., social network benchmark)\cite{10.1145/2723372.2742786}, and an RDF benchmark (i.e., semantic publishing benchmark) targeted for industrial systems following a real-life application scenario of British Broadcasting Corporation (BBC) semantic publishing platform.

BigDataBench is a recent joint effort by several academic and industrial partners for construction of standard bigdata benchmark suite which includes graph analysis workloads as part of its workloads \cite{DBLP:journals/corr/WangZLZYHGJSZZLZLQ14}. Most of the details of the BigDataBench has been described in Section \ref{sec:bigdata-benchmarks-with-graph-data-processing-capabilities} previously.

GraphChallenge is a graph based performance benchmarking effort which uses the knowledge learned from previous efforts such as GraphAnalysis, Graph500, etc. \cite{website:graphchallenge:2018a}. It creates a new set of challenges to advance the community. We summarize the key observations made in our study in next section.

\begin{figure}[htbp]
	\centering
		\includegraphics[width=0.7\textwidth]{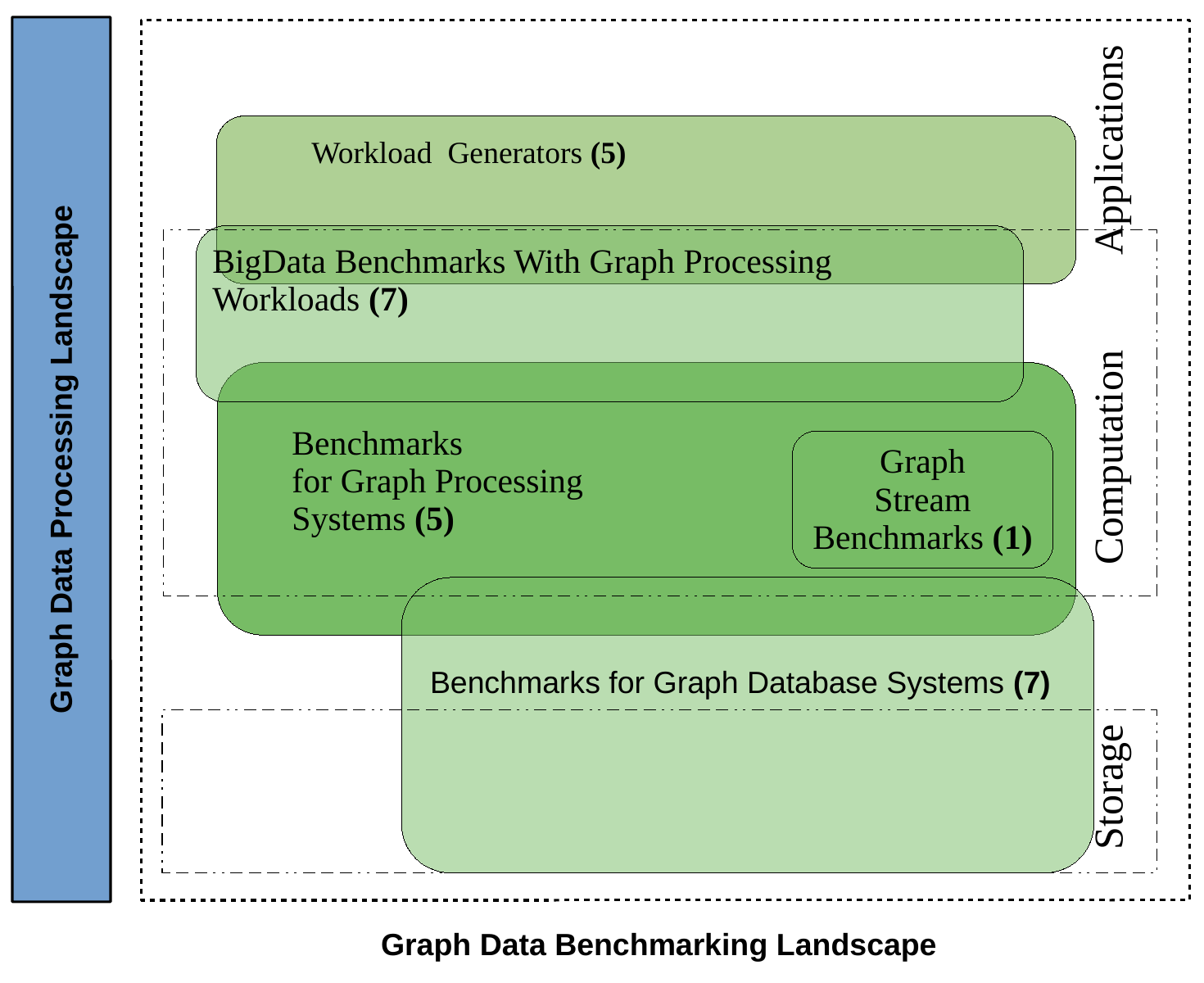}
	\caption{Summary view of graph benchmarking landscape.}
	\label{fig:graphbenchmarkinglandscape}
\end{figure}

\section{Key Observations}
\label{sec:comparison}
Overall this paper did a detailed investigation of the benchmarks used for performance analysis of graph data management and processing systems.

Out of the thirteen benchmarks (see Figure \ref{fig:graphbenchmarkinglandscape} for breakdown. The values within brackets indicate the number of benchmarks.) developed specifically for graph data management and processing only few (such as HPCSGAB) use graph specific metrics such as TEPS. Latency related metrics were used by most of the benchmarks. Latency is relatively easily measured compared to metrics such as TEPS. Different graph processing systems require different levels of improvements (i.e., instrumentation) to accustom to use advanced metrics such as TEPS. Although more than a decade has passed since the introduction of TEPS, our survey indicates still the graph processing community at large are facing practical difficulties in adopting it in their benchmarks.

\begin{figure*}[htbp]
	\centering
		\includegraphics[width=1\textwidth]{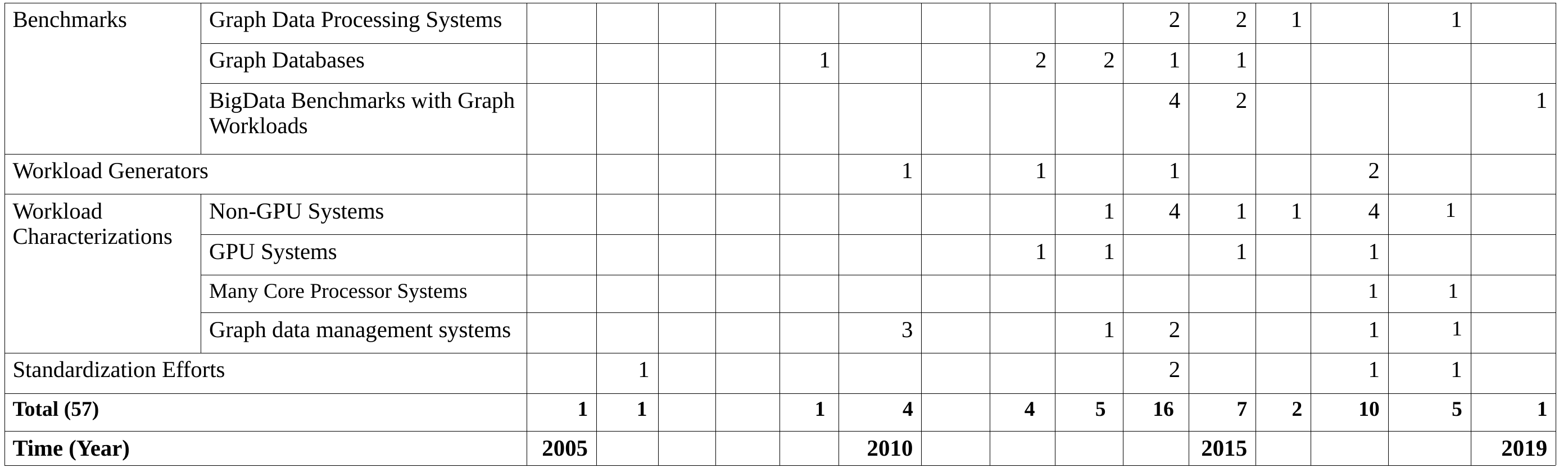}
	\caption{Performance Analysis of Graph Data Management and Processing Systems. The timeline gives a briefing on the key areas investigated.}
	\label{fig:timeline}
\end{figure*}

Out of the 58 works surveyed in this paper 45\% appeared in last 5 years time period (See Figure \ref{fig:timeline}). Half of the works (29 in total) appeared in 2010-2014 time frame (5 years). This is a significant improvement compared to the 2005-2009 (5 years) time frame. We expect this trend to continue in the near future as well.

The survey indicated that most of the cutting edge graph processing benchmarks do not support emerging areas such as benchmarking streaming graph data processing systems and graph based machine learning. However, data stream processing has emerged as a key technological drive for business value creation and there is an increasing attention from the industry for deploying applications of stream data processing. Similarly increasing number of work have been done on running machine learning on graph data sets. But relatively few benchmarks currently exist for these area.

Each and every summary table corresponding to benchmarks we have listed the main programming languages used for developing them. Out of the benchmarks developed for graph databases (7) as well as bigdata benchmarks with graph data processing workloads (7) nine benchmarks has been developed using Java (64\%). This indicates that Java as a programming language had significant impact on the graph data processing benchmark development. However, the usefulness of Java based benchmarks for large scale workload scenarios is questionable. Many of such large scale benchmarks have been developed in C/C++ languages (E.g., Graph 500). There have been other alternatives such as X10 being investigated \cite{Dayarathna2017}. Next section concludes the paper.

\section{Summary and Future Research Directions}
\label{sec:summary-future}
This paper made a detailed survey of benchmarks for graph data management and processing systems. It mainly investigated on benchmarks for graph data management and processing systems, workload generators, workload characterizations, standardization efforts, etc. A summary of the research described in this paper (15 years in total) is shown in Figure \ref{fig:timeline}. The number in each cell corresponds to the number of research works done in that particular year. Significant increase of the number of works done on performance analysis of storing and processing of graphs are found during the last 8 years (2012-2019) time period which amounts to 88\% out of the 57 listed in Figure \ref{fig:timeline}.

Graph processing needs to conduct benchmarking with realistic data. Real users (i.e., system owners) have different information to handle compared to the network characteristics shown by synthetic datasets. Real graphs are mostly property graphs. They have real properties and temporal characteristics. Hence graph benchmark developers need to concentrate on generating realistic property graphs. 

Another area of significant importance for future investigations is the design and implementation of graph stream benchmarking. While there have been multiple graph stream algorithms being developed hardly any work been done on investigating the performance aspects of such systems in processing graph stream algorithms \cite{McGregor:2014:GSA:2627692.2627694}.

There is an increasing interest on running machine learning (e.g., deep learning) on graph data which is emerging as a new area of graph processing \cite{DBLP:journals/debu/HamiltonYL17}. Another area of interest would be conducting privacy preserving computation on top of graph data \cite{Nayak:2015:GPS:2867539.2867661}. However, the present research do not concentrate on these areas. We envision there will be increasing number of work appear on investigating and improving the performance of such applications in near future.

Novel performance metrics related to system performance has been investigated by graph processing community such as power consumption of graph processing systems \cite{7363775}\cite{Khan:2016:EEL:2968219.2968296}. There are other factors such as fault tolerance, energy saving, and security guarantee to be considered when designing graph management and processing benchmarks \cite{10.1007/978-3-642-41278-3_75}.

\bibliographystyle{unsrt}  


\end{document}